\documentclass[iop]{emulateapj}

\newcommand{\hone}{H~{\footnotesize{I}}}  	
\newcommand{\heone}{He~{\footnotesize{I}}}  	
\newcommand{\catwo}{Ca~{\footnotesize{II}}}  	
\newcommand{\cartwostar}{C~{\footnotesize{II}}$^{*}$}  	
\newcommand{\carfour}{C~{\footnotesize{IV}}}  	
\newcommand{\naone}{Na~{\footnotesize{I}}}	
\newcommand{\nitfive}{N~{\footnotesize{V}}}  	
\newcommand{\oxyone}{O~{\footnotesize{I}}}  	
\newcommand{\oxysix}{O~{\footnotesize{VI}}}  	
\newcommand{\oxyseven}{O~{\footnotesize{VII}}}  
\newcommand{\oxyeight}{O~{\footnotesize{VIII}}} 
\newcommand{\dirbe}{{{DIRBE}}}		
\newcommand{\fimsspear}{{FIMS/SPEAR}}	
\newcommand{\fuse}{{\it{FUSE}}}			
\newcommand{\iras}{{\it{IRAS}}}			
\newcommand{\xmmnewton}{{\it{XMM-Newton}}}	
\newcommand{\rosat}{{\it{ROSAT}}}		

\shorttitle{Draco Halo}
\shortauthors{Shelton et al.}

\begin{document}

\title{Hot Gas in the Galactic Thick Disk and Halo Near the Draco Cloud}

\author{R. L. Shelton\altaffilmark{1},
	D. B. Henley\altaffilmark{1}, and
	W. V. Dixon\altaffilmark{2} }
\altaffiltext{1}{Department of Physics and Astronomy, the University of Georgia, Athens, GA 30602}
\altaffiltext{2}{Department of Physics and Astronomy, the Johns Hopkins 
University, Baltimore, MD 21218}
\email{rls@physast.uga.edu}

\begin{abstract}

This paper examines the ultraviolet and X-ray photons generated
by hot gas in the Galactic thick disk or halo in the Draco region
of the northern hemisphere.    Our analysis uses the intensities 
from four ions, \carfour, \oxysix, \oxyseven, and \oxyeight,
sampling temperatures of $\sim10^5$ to 
$\sim3 \times 10^6$~K.   
We measured the \oxysix, 
\oxyseven, 
and \oxyeight\ 
intensities from 
\fuse\ and \xmmnewton\ data and subtracted off the 
local contributions
in order to deduce the thick disk/halo contributions.
These were supplemented with published \carfour\ 
intensity and 
\oxysix\ column density measurements.
Our estimate of the thermal pressure in the \oxysix-rich
thick disk/halo gas, 
$p_{th}/k = 6500^{+2500}_{-2600}$~K~cm$^{-3}$, 
suggests that the thick disk/halo is more highly pressurized than 
would be expected from
theoretical analyses.
The ratios of \carfour\ to \oxysix\ to \oxyseven\ to \oxyeight\
intensities were compared with those predicted by 
theoretical models.
Gas which was heated to $3 \times 10^6$~K
then allowed to cool radiatively
cannot produce enough
\carfour\ or \oxysix-generated photons
per \oxyseven\ or \oxyeight-generated photon.
Producing enough \carfour\ and \oxysix\ emission requires 
heating additional gas to $10^5 < T < 10^6$~K.
However, 
shock heating, which provides heating 
across this temperature range, overproduces \oxysix\ relative
to the others.
Obtaining the observed mix may require 
a combination of several processes, 
including some amount of shock heating, heat conduction, 
and mixing, as well as radiative cooling of very hot gas.

\end{abstract}


\section{Introduction}

Observations of \carfour, \oxysix, \oxyseven, and \oxyeight\
show that hot gas 
extends a few kpc above the plane 
\citep{savage_wakker_09,yao_etal_09}
into a region historically called the halo and 
more recently called the thick disk. 
This is an active and dynamically important
region of the Galaxy.
On-going star formation has been seen in the
thick disks of other spiral galaxies
\citep{rueff_howk_10}, implying that these thick disks
continue to experience the winds and
supernova (SN) explosions that result from young stars.   
In addition, thick disks churn as 
outflows from the thin disk, long postulated theoretically
\citep{shapiro_field_76}
and more recently seen in numerical simulations
having normal SN rates
\citep{joung_maclow_06,deavillez_breitschwerdt_07},
make their way into 
the thick disk.
Estimates based on the observed \carfour, \oxysix,
1/4~keV, and 3/4~keV intensities at high southern latitudes
indicate a large radiative loss rate;
the region
above the thin disk is so luminous 
that it radiates 2/3 of the Galaxy's supernova power 
\citep{shelton_etal_07}.
Since most of the Galaxy's SN energy 
is injected nearer to the
midplane, a large 
luminosity produced above the thin disk
suggests the
sort of churning seen in simulations.
Not only is the thick disk affected by {\it{in situ}} heating and
upwellings from 
the thin disk beneath it, but it is also affected by 
inflows from above, as in the case of high velocity
clouds colliding with the Galaxy \citep{tripp_etal_03}, 
or if we consider
the more extended halo, in the form of accretion from
the intergalactic medium 
(e.g., \citealt{rasmussen_etal_09}).

Most of the volume within a few kpc of the midplane is
filled with very hot gas 
having temperatures in excess of $10^6$~K 
(see pathlength estimates in 
\citet{shelton_etal_07} or scale height estimates in 
\citealt{yao_wang_07}).   This gas is traced by 
\oxyseven\ ions 
and 
\oxyeight\ ions 
having collisional ionization equilibrium (CIE) temperatures,
$T_{CIE}$, of $\sim 1 \times 10^6$~K and 
$\sim 3 \times10^6$~K, respectively.
Somewhat cooler, but still hot, gas
is traced by \carfour\ ions 
($T_{CIE} \sim 1 \times 10^5$~K) 
and \oxysix\ ions
($T_{CIE} \sim 3 \times 10^5$~K).
It fills a lesser fraction
of the space (see pathlengths in \citealt{shelton_etal_07})
but accounts for more radiative energy loss than
does the hotter gas.   Because gas in this temperature regime
cools rapidly, it must be
replenished 
from a nearby reservoir of
hotter gas.    For this reason, a common
conception of the hot interstellar medium (ISM) is one in which
1 to $3 \times 10^5$~K gas resides in transition zones
between hotter and cooler gas.   Some authors
(e.g., \citealt{savage_wakker_09}) have begun to call
the 1 to $3 \times 10^5$~K material ``transition temperature''
gas.   Here, for simplicity, however, 
we use the
term ``hot gas'' for the entire $\sim 1 \times 10^5$ to
$\sim 3 \times 10^6$~K regime.

The tracers of hot gas are observed by
UV and X-ray instruments.
Ultraviolet instruments are used to observe the 
strong resonance line transitions 
(2s $^2$S$_{1/2}$ - 2p $^2$P$_{3/2}$ and
2s $^2$S$_{1/2}$ - 2p $^2$P$_{1/2}$) of \carfour\ and \oxysix.
Owing to the excellent spectral resolution
of recent
UV instruments, interstellar \carfour\ and \oxysix\
have been studied well via absorption
line spectroscopy \citep{savage_etal_03,bowen_etal_08,savage_wakker_09}.   
They have also been seen via emission
spectroscopy, but 
along fewer sight lines
(e.g., \citealt{shelton_etal_01,dixon_sankrit_08,park_etal_09}). 
In contrast, X-ray instruments are used to observe 
K$\alpha$ transitions of 
\oxyseven\ and \oxyeight.
Due to the limited spectral resolution and/or throughput of 
current 
X-ray instruments, it is far easier to detect photons
produced by interstellar \oxyseven\ and \oxyeight\ ions than to 
identify \oxyseven\ and \oxyeight\ absorption profiles
in the spectra of distant X-ray point sources.  
While some
absorption column densities
have been measured
(e.g., \citealt{yao_wang_07,bregman_lloyddavies_07}),
most of the \oxyseven\ and \oxyeight\ measurements
are of emitted photons (e.g., \citealt{henley_shelton_10}).

Column density and emission intensity measurements
plumb different aspects of the interstellar gas.
The column density
depends only on the path length and the density of pertinent
ions in the gas.    
The emission intensity, however, depends on 
two additional factors,
the electron density and the emissivity, which is
a function of temperature.   Thus, high ions that
are much cooler than their CIE
temperature (for example, ions resulting from
photoionization and ions that have cooled in a 
non-equilibrium fashion)
can be seen via absorption spectroscopy, but 
are very inefficient radiators.
This is especially 
relevant for \carfour, because a significant
fraction of the \carfour\ ions seen in absorption measurements
are photoionized \citep{savage_wakker_09}.

Here, we focus on the emission intensities from
\carfour, \oxysix, \oxyseven, and \oxyeight\ 
from a single region of the sky, 
($\ell = 90\degr, b \sim 40\degr$), a
region near, but not toward, the Draco molecular cloud complex
(MBM 41 through 44, \citealt{magnani_etal_85}).   
This paper is organized as follows.
In Section~\ref{sect:data},
we assemble measurements of \carfour, \oxysix,
\oxyseven, and \oxyeight.   
The \carfour\ intensity measurements presented
in Subsection~\ref{sect:civ} come from published 
observations made by \fimsspear.
We complement these with new measurements of
the \oxysix\ intensity, made from \fuse\ data
(Subsection~\ref{sect:ovi}).  
In Subsection~\ref{sect:NOVI_estimate}, we deduce
the extraplanar \oxysix\ column 
density in the Draco region from existing \fuse\
measurements.
In Subsection~\ref{ovii_ovii_estimate}, we present our
\oxyseven\ and \oxyeight\ measurements, made from
archival \xmmnewton\ observations of a direction 
$1\degr$ from  our \fuse\ and \fimsspear\ direction.
For each ion, we
take care to remove the local contributions (from 
the Local Bubble and/or heliosphere) and
compensate for interstellar absorption.  The results are
the second known foursome of intrinsic \carfour, \oxysix, \oxyseven,
and \oxyeight\ intensities for the region of the 
thick disk/halo that is above the thin disk and
toward a single part of the sky,
and the first such foursome for the northern Galactic hemisphere.
In Section~\ref{sect:discussion}, we use the 
\oxysix\ emission intensity and absorption column density
to determine the
density and pressure of the \oxysix-rich gas in the
thick disk/halo above the thin disk.   
We then use the
ratios of our \carfour, \oxysix, \oxyseven, and \oxyeight\
intensities to test phenomenological models.
These include a radiative
cooling model, such as would be expected for 
accretion of intergalactic gas onto the galaxy,
a point-injection model,
such as would be expected for
supernova explosions,
and a turbulent mixing model.
We find that
radiative cooling of very hot gas is not able to
produce enough \carfour\ and \oxysix\ photons per
\oxyseven\ or \oxyeight\ photon.
Bubbles formed from point injections of $5 \times 10^{50}$~
ergs have the opposite problem.
Models of turbulent mixing between very hot gas and cooler gas
do not provide \oxyseven\ or \oxyeight\ predictions,
but can explain high ratios of \carfour\ to \oxysix\
emission.
We also present
the ratio of \carfour, \oxysix, \oxyseven, and \oxyeight\
intensities, so that they may be compared with other
model predictions as they become available.
In Section~\ref{sect:summary},
we summarize our results, concluding that 
energy must have been injected into the halo
in a way that heated some of the gas to 
$\sim3 \times 10^6$~K
while heating other gas to 1 to $3 \times 10^5$~K.

\section{Data} 
\label{sect:data}

\subsection{The Halo \carfour\ Intensity}
\label{sect:civ}

As part of a survey program, 
the Draco region was observed by \fimsspear\
\citep{edelstein_etal_06b},
which is sensitive to \carfour\ resonance line
doublet photons 
($\lambda = 1548, 1551$~\AA), between
February and May of 2004.
\citet{park_etal_09} presented measurements of the 
doublet's intensity 
as a function of location within the Draco neighborhood 
in the form of a smoothed, pixelated map.
The direction of interest, 
$\ell = 90.0\degr, b = 39.6\degr$
is located near the intersection of four
0.1$\degr$-wide pixels in 
their map.
For these four pixels, 
the average intensity is 
$\sim5500$ in units of ph~s$^{-1}$~cm$^{-2}$~sr$^{-1}$,
hereafter line units (LU), and the range of intensities
is 700 LU.
Because \citet{park_etal_09} reported a signal to noise ratio
of $>3.0$ for intensities over 5000 LU, we
take $1\sigma$ to be 1/3 of the measured intensity
and adopt 
$I$ = $5500 \pm 1830$ LU
as the observed intensity in the doublet.
This and later measurements are tabulated in 
Table~\ref{table:allintensities}.
This value
is very near to the average intensity reported by
\citet{park_etal_09} for all off-cloud
regions within their map: $5559 \pm 1008$ LU.

\begin{deluxetable}{lccc}
\tablewidth{0pt}
\tablecaption{Observed and Extraplanar Intensities of \carfour,
\oxysix, \oxyseven, and \oxyeight}
\tablehead{
\colhead{ Ion }     
& \colhead{ $\ell,b$ }
& \colhead{ Observed }
& \colhead{ Intrinsic Extraplanar  } \\
\colhead{  }     
& \colhead{ }
& \colhead{ Intensity }
& \colhead{ Intensity } \\
\colhead{  }     
& \colhead{ Degrees }
& \colhead{(ph s$^{-1}$ cm$^{-2}$ sr$^{-1}$)}
& \colhead{(ph s$^{-1}$ cm$^{-2}$ sr$^{-1}$)}
} 
\startdata
\carfour\ & $90.0\degr, 39.6\degr$ & 5500 $\pm$ 1830 & 6540 $\pm$ 2180 \\ 
\oxysix\ & $90.0\degr, 39.6\degr$ & 3500 $\pm$ 940 & 4770$^{+1300}_{-1380}$\\
%
\oxyseven\ & $90.0\degr, 38.4\degr$ & $6.8^{+0.9}_{-0.6}$ & $6.4^{+1.4}_{-1.5}$ \\
\oxyeight\ & $90.0\degr, 38.4\degr$ & $1.0^{+0.4}_{-0.3}$ & $1.0 \pm 1.2$\\
\enddata
\tablecomments{ The extraplanar \oxysix\ column density in
this region is
$N_{OVI} = 1.9 \pm 0.5 \times10^{14}$~cm$^{-2}$. \\
The \carfour\ and \oxysix\ intensities are for the doublets.
}
\label{table:allintensities}
\end{deluxetable}

No observations of foreground \carfour\ emission have 
been reported in the literature and little emission is expected
from the Local Bubble or heliosphere.
Therefore, we assume that the
\carfour-rich gas resides above the Local Bubble, and also 
above the Galaxy's
\hone\ and/or dust-rich layer.
The photons emitted in this region would be subject to
extinction by intervening gas.
The degree of obscuration determined from the \hone\ 
column density
from the Leiden-Argentine-Bonn (LAB) survey 
of neutral hydrogen 
\citep{kalberla_etal_05}
is roughly similar to that determined from
the \dirbe-corrected \iras\
data \citep{schlegel_etal_98}.
The LAB survey found  
a column density of 
 $N_{HI} = 1.52 \times 10^{20}$~cm$^{-2}$,
which implies a loss of 
$20\%$ 
of the intensity, 
based on the color excess to \hone\ column density relation empirically
determined by \citet{diplas_savage_94}
and the extinction relation calculated by \citet{fitzpatrick_99}.
The \dirbe-corrected \iras\ 100 $\mu$m intensity is
$I_{100} = 0.71$~MJy~sr$^{-1}$,
which implies a loss of 
$11\%$, 
based on the $I_{100}$ to color
excess conversion used by \citet{schlegel_etal_98} and the 
extinction relation calculated by \citet{fitzpatrick_99}.
Thus,  if all of the obscuring material  lies between the source of the 
photons and the Earth,
then the halo's intrinsic intensity would be
$6200\pm2070$ to $6880\pm2290$ LU,
assuming the $11\%$ and $22\%$ loss rate, respectively.
Here, we take the halo's intrinsic \carfour\ doublet intensity
to be the average: 
$6540\pm2180$ LU.
Note that throughout the paper, the unrounded measurement
values are used in the calculations, but the rounded results
are reported.

%
%

\subsection{The Halo \oxysix\ Intensity}
\label{sect:ovi}
\subsubsection{\fuse\ Observations and Data Reduction}

We obtained two \fuse\ observations for this project.
The first, observation ID E12401,
was directed toward the Draco 
cloud, $\ell = 89.7\degr, b = 38.5\degr$.
Beginning on 2005-11-30, six exposures were taken 
for a total of 5,144 seconds. 
The \fuse\ spacecraft failed before the remainder of
the observation could be completed.  
The second observation, observation ID E12402, 
was directed toward an unobscured region
near the Draco cloud,
$\ell = 90.0\degr, b = 39.6\degr$.
Several exposures began on 2006-11-15 and the remainder
began on 2007-3-14.
The total exposure time was 79,912 seconds, of which
31,316 seconds were taken at night.
The original intent of this program was to perform a 
cloud shadowing
study.    However, the on-cloud observation was too short 
to yield clear detections 
(see subsection \ref{subsect:oviintensity}).     
Here, we focus on the off-cloud observation.




We examined the data taken through the largest
aperture of the LiF1A channel,
as this yields spectra with the greatest signal to noise ratio.
Like other observations of diffuse emission, the data 
were taken in time-tag mode.   
Photon lists for each exposure were processed with version 3.2 of the
CalFUSE pipeline.  We used the extended source extraction window
option, disabled the background subtraction feature, and set the
pulse height limits to 2 and 25 for this processing.

Earlier analyses found that the wavelength scale must be re-calibrated
by comparisons with emission lines of known wavelengths 
\citep{shelton_etal_01}.   Because each exposure may have a different
offset, the exposures were processed by CalFUSE and re-calibrated
before being added.    The wavelength offsets were found by 
aligning the observed Lyman $\beta$
airglow feature in the LiF1A spectra with the theoretical
Lyman $\beta$ wavelength at zero velocity in the geocentric 
reference frame.    After being corrected in this manner,
the wavelength scale for each spectrum was converted to the 
heliospheric reference frame and the spectra added together.

\subsubsection{\oxysix\ Intensity Measurements}
\label{subsect:oviintensity}

\begin{figure}
\epsscale{0.8}
\plotone{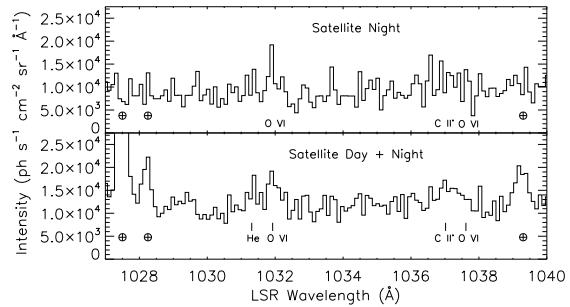}
\vspace{1cm}
  \caption{LiF 1A spectra of the off-cloud observation.
The spectra have been converted to the Local Standard of Rest
reference frame and binned into 0.104 \AA\ pixels.
The rest wavelengths of the \oxysix, \cartwostar, 
2nd order \heone, and Earth
airglow features ($\oplus$) are marked on the plots in the
LSR reference frame.
Top:  Data taken during the night portion 
of the satellite's orbit.
Bottom:   Data taken during both the night and day portions of the
orbit.  During the orbital day, the detectors are 
subjected to 
\hone\ and \oxyone\ airglow photons and also record second-order 
diffraction counts from \heone, which appear around 1031 \AA.
The rest wavelengths of the \oxysix\ and \cartwostar\ emission
lines are marked with vertical bars on the lower plot.}
\label{fig:myspectra}
\end{figure}

Figure~\ref{fig:myspectra} 
shows the LiF1A spectrum in the 1027 to 1040 \AA\ region.
The \oxysix\ 1032 \AA\ resonance line is apparent in both the
night-only and the day+night spectra.   The \oxysix\
1038~\AA\ resonance line 
is naturally dimmer and also easily obscured
by the neighboring \cartwostar\ 1037 \AA \ feature.
The bright peaks near 1027.5, 1028, and 1039 \AA\ in the
day+night spectrum are due to atmospheric hydrogen and
oxygen.   The 1031 \AA\ feature that appears in the
day+night spectrum, but not in the night-only spectrum, is
thought to be the second order diffraction line of atmospheric 
\heone\
\citep{shelton_etal_07}.

\begin{figure}
\epsscale{1.0}
\plottwo{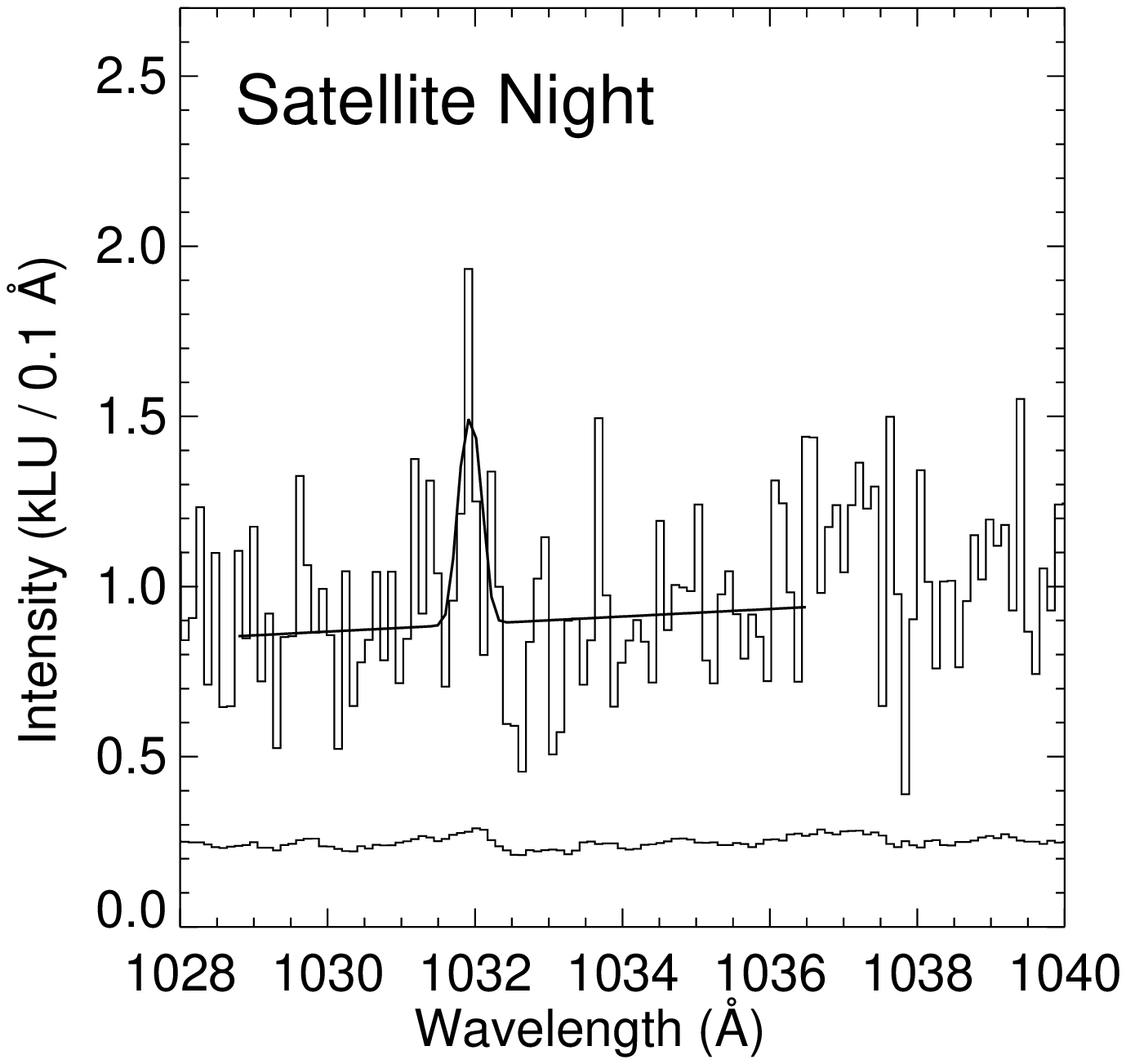}{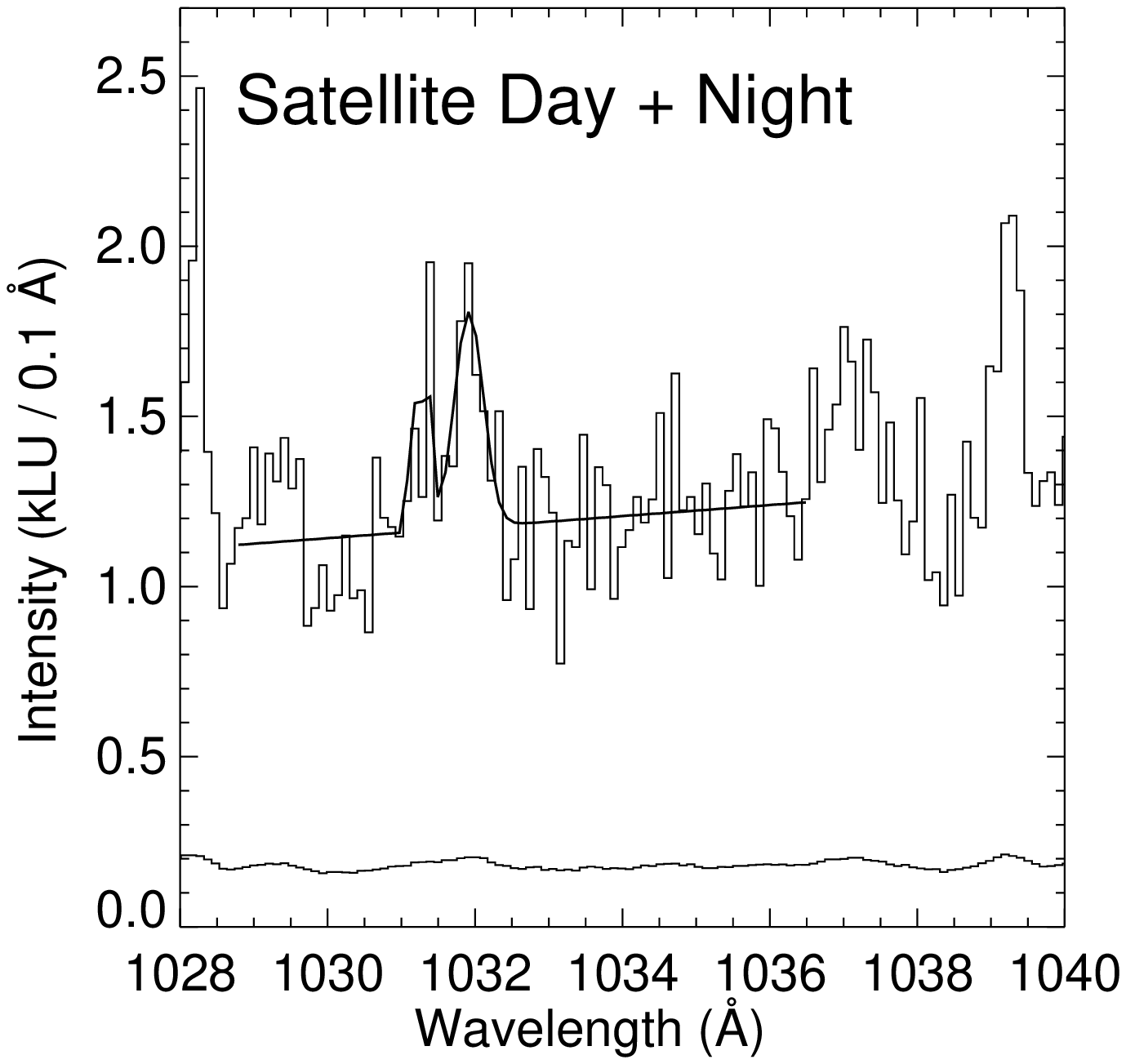}
%
  \caption{  Method $\#2$ fits to the spectra (left:  night-only data;
right:  day+night data).
The plotted spectra have been converted to the LSR reference frame and binned 
into 0.104 \AA\ pixels.   
The Method $\#2$ fitting functions are overplotted.   They consist of 
a linear function that models the continuum, a Gaussian convolved with
a tophat that models the 1032 \AA\ feature, and in the case of the
day+night data, a second Gaussian convolved with
a tophat that models the airglow 1031 \AA\ feature. 
The ragged curves near the bottom of each plot are the error bars on the 
data. }
\label{fig:vansplot}
\end{figure}

Here, we focus on the \oxysix\ 1032 \AA \ line in
the off-cloud data.
We measure its strength, both by the method of searching
for significant excesses in counts that was employed in 
\citet{shelton_etal_01}
and later papers (method $\#1$) and 
by 
a modified version of 
the method of fitting with a Gaussian function 
that has been
convolved with a 106 km sec$^{-1}$ tophat function 
that was employed in
\citet{dixon_etal_01}
and later papers
(method $\#2$).   
The modification is to add a second tophat-convolved
Gaussian to model the second order \heone\ airglow feature that 
appears at 1031~\AA\ in the daytime data.
See Figure~\ref{fig:vansplot} for the fitting results.
Both methods cleanly separate the
\oxysix\ 1032 \AA\ feature from the  daytime 1031~\AA\ feature.
Table~\ref{table:intensityresults} presents the intensity measurements.     
%
%
We average the measurements, yielding an observed intensity
in the \oxysix\ 1032~\AA\ line of  $2340 \pm 630$~LU.

\begin{deluxetable*}{lcccc}
\tablewidth{0pt}
\tablecaption{Observed Intensities and 1 $\sigma$ Statistical Uncertainties
for the 1032 \AA\  \oxysix\ feature}
\tablehead{
\colhead{  }     
& \colhead{Night Only }
& \colhead{Night Only }
& \colhead{Day+Night  }
& \colhead{Day+Night  } \\
\colhead{  }     
& \colhead{Method $\#$1}
& \colhead{Method $\#$2}
& \colhead{Method $\#$1} 
& \colhead{Method $\#$2} \\
\colhead{  }     
& \colhead{(ph s$^{-1}$ cm$^{-2}$ sr$^{-1}$)}
& \colhead{(ph s$^{-1}$ cm$^{-2}$ sr$^{-1}$)}
& \colhead{(ph s$^{-1}$ cm$^{-2}$ sr$^{-1}$)}
& \colhead{(ph s$^{-1}$ cm$^{-2}$ sr$^{-1}$)}
} 
\startdata
off-cloud: &  1970 $\pm$ 600 &  2240 $\pm$ 660 &  2140 $\pm$ 470 & 2990 $\pm$ 790 \\ 
%
on-cloud: & &0 $\pm$ 690 &-240 $\pm$ 1600 & 0 $\pm$ 1030\\
\enddata
\tablecomments{
For the on-cloud data,
method $\#1$ was unable to fit the continuum for the night-only data. 
Method $\#2$ calculated the 
3$\sigma$ upper limit.
For comparison with the other values, the 3$\sigma$ upper limits found
by method $\#2$ were
converted into a ``measurement $\pm 1\sigma$'' form by dividing by 3.}
\label{table:intensityresults}
\end{deluxetable*}

To place the \oxysix\ 1032 \AA\ 
intensity along our off-cloud sight line into 
context, we compare it with \oxysix\ 1032 \AA\ intensities 
measured for other 
high-latitude lines of sight.  Using archival \fuse\ data, 
\citet{dixon_etal_06} and \citet{dixon_sankrit_08} have searched 
for diffuse \oxysix\ photons along roughly 300 lines of sight. 
Here we consider the 54 sight lines in their sample (plus E12402) 
with Galactic latitudes of $|b| > 20\degr$ and orbital-night
exposure times exceeding 18~ksec.  This sample is complete in the 
sense that any emission 
brighter than 2000 LU would have been detected in an 18~ksec 
night-time exposure.  Within this sample, 63$\%$ 
have statistically-significant (i.e., $> 2\sigma$) intensities 
of \oxysix\ 
1032 \AA\ photons, the greatest of which is 5500 LU.    If null 
detections
are treated as having zero flux and a 1~$\sigma$ error, then the
median and mean intensities in the 1032 \AA\ line are
2100 and 1800~LU, respectively,
while the standard deviation 
about the mean
is 1620~LU.
Our off-cloud sight line, with an observed 1032 \AA\ intensity of 
$2340\pm 630$~LU,
lies within one sigma of the mean, so it is not at all unusual.

Our measurements of the weaker \oxysix\ 1038 \AA\ feature 
were not as
sound as our measurements of the stronger 1032 \AA\ feature.     
The 1038 \AA\ feature sits to the right of the 
1037 \cartwostar\ line
and to the left of the \oxyone\ airglow feature, making it 
impossible to determine the background continuum accurately.
For this reason, we present neither day$+$night nor night-only
1038 \AA\ measurements.

We do, however, present measurements from our search for a 
second \oxysix\ 1032 \AA\ feature.    This search was motivated
by the observation of a weak intermediate velocity 
($V_{LSR} = -112$ km sec$^{-1}$)
feature in the LAB survey 
of galactic \hone.     If infalling gas at this LSR velocity was accompanied
by \oxysix, then the stronger line in its doublet would appear at a wavelength of 
1031.49 \AA\ in our heliocentric restframe spectra.    We searched
for such a feature in a 0.36 \AA\ wide window (corresponding to the
width of the LWRS aperture) in the night-only data.    We did not
search for the feature in the day+night data, because it would have 
overlapped with the second order \heone\ feature.    Using method $\#1$,
we found a statistically insignificant excess intensity  of
210 $\pm$ 480~LU.

We examined the on-cloud data, as well.     These spectra were of
very poor quality, because of the short observation time
(5144 seconds of day$+$night data, including 2056 seconds
of night-only data.) 
Neither method $\#1$ nor method $\#2$ could find emission
features around 1032 or 1038 \AA.     
We were able to use method $\#1$ to determine and subtract the continuum and
then measure the residual intensity within the wavelength range where the
1032 \AA\ feature was expected to be found, based on the location
of the 1032 \AA\ feature in 
the off-cloud spectra taken from the day$+$night data.    
This method yielded a measured  intensity of 
-243 $\pm$ 1604~LU, 
where the negative sign indicates that this region of the spectrum
has fewer counts than the fitted continuum.
Method $\#2$ yielded 3 $\sigma$ upper limits of 
3090 and 2060~LU 
from the day$+$night and night-only data, respectively.   In
both cases, the uncertainties are too large 
to constrain our conceptions of the interstellar
medium.

In order to determine the velocities using method $\#1$, the heliocentric
spectra were refit using a Gaussian for the feature in question and a low-order
polynomial for the continuum.    
Method $\#2$ also yielded velocity estimates. 
In both cases, the reference frames of the
analyzed spectra were heliospheric.   The resulting velocities
have been converted to the LSR reference frame by the addition
of  16.73 km sec$^{-1}$. 
(See Table~\ref{table:velocityresults} for the measurement results.)
The wavelength scale is accurate to 
$\sim10$~km~sec$^{-1}$ \citep{shelton_etal_01}.

\begin{deluxetable}{lcccc}
\tablewidth{0pt}
\tablecaption{Velocity with Respect to the Local Standard of Rest}
\tablehead{
\colhead{  } 
& \colhead{Night Only }
& \colhead{Night Only }
& \colhead{Day+Night  }
& \colhead{Day+Night  } \\
\colhead{  }     
& \colhead{Method $\#$1}
& \colhead{Method $\#$2}
& \colhead{Method $\#$1}
& \colhead{Method $\#$2} \\
\colhead{  }     
& \colhead{(km sec$^{-1}$)} 
&\colhead{(km sec$^{-1}$)} 
& \colhead{(km sec$^{-1}$)}
& \colhead{(km sec$^{-1}$)}
}
\startdata
\oxysix\ 1032 \AA   & $-12 \pm 7$  & $2\pm17$  &  $-9 \pm 11$  & $-3\pm14$ \\

\enddata
\label{table:velocityresults}
\end{deluxetable}

\subsubsection{Intrinsic Halo \oxysix\ Intensity}
\label{subsect:intrinsic}

For optically thin plasmas, the 1038 \AA\ line is half as
strong as the 1032 \AA\ line.   Assuming this ratio, the
doublet intensity becomes 
$3500 \pm 940$~LU 
(Table~\ref{table:allintensities}).
The observed intensity is due to photons emitted in the halo
but attenuated by intervening material plus photons emitted
locally, by the Local Bubble if it exists.
%
%
%
When the observations were planned, it was intended to use
the on-cloud measurement to estimate the foreground
intensity.    However, because the exposure time was too 
short for a good measurement, 
we take the local
contribution from the only other
\oxysix\ cloud shadow observation, that toward 
$\ell=278.6\degr,b=-45.3\degr$ where a cloud lies
$230\pm30$~pc from Earth.
That cloud 
observation yielded upper limits on the \oxysix\ 1032 \AA\
and 1038 \AA\ line intensities.   The tightest upper limit
is that on the 1032 \AA\ line in the day$+$night data.
Multiplying it by 1.5 in order to account for the 1038 \AA\
contribution yields a limit on the doublet intensity of
$30^{+340}_{-30}$~LU
\citep{shelton_03}.    
The unrounded values were made
available by the author.  
Subtracting them from our unrounded doublet intensity yields 
$3480^{+940}_{-1000}$~LU.

If the non-local photons originated
above even part of the Galaxy's \hone\ layer, then the 
intrinsic intensity would be greater than 
$3480^{+940}_{-1000}$~LU.
Here we make the assumption that the
\oxysix-rich gas resides above the full extent of the Galaxy's
\hone\ and/or dust-rich layer.
As with the \carfour\ intensity, we estimate the degree of 
obscuration of the halo \oxysix\ emission using the 
LAB \hone\ column density \citep{kalberla_etal_05}
and the \dirbe-corrected \iras\ 100 $\mu$m intensity 
\citep{schlegel_etal_98}.
Using the same input values and conversion relations that 
we used in Section 2.1, we find that the LAB \hone\ 
column density and the 100 $\mu$m intensity imply \oxysix\ 
losses of 
33$\%$ and 19$\%$, respectively. 
%
Thus,  if all of the obscuring material  lies between the source of the 
photons and the Local Bubble,
then the halo's intrinsic intensity would be
$4320^{+1170}_{-1240}$ to $5220^{+1420}_{-1510}$~LU.
Here, we take the halo's intrinsic \oxysix\ doublet intensity
to be the average: 
$4770^{+1300}_{-1380}$~LU,
which is also listed in Table~\ref{table:allintensities}.
%
%

The optical thickness of the on-cloud line of sight is much larger
($I_{100} = 3.48$ MJy~sr$^{-1}$,  implying a loss of 
70$\%$ 
of the incoming photons).    If the gas above the cloud produces
the same doublet intensity as 
that derived above for the off-cloud line of sight, then only
$1440^{+390}_{-420}$~LU
would survive passage
through the cloud and reach the detector.   Of this, 
$960^{+260}_{-280}$~LU
would be in the \oxysix\ 1032 \AA\ line.
This intensity is within our
observational upper limits.

\subsection{An Estimate of the Halo's \oxysix\ Column 
Density}
\label{sect:NOVI_estimate}

We estimate the line of sight 
\oxysix\ column density in nearly the same manner
as \citet{shelton_etal_07}, namely by averaging the 
column densities on the four nearest directions for
which they are measured and calculating the uncertainty from the
typical scatter in measured values.  
The difference between our method and that of 
\citet{shelton_etal_07} is that we 
also include the measurement error; see below.
We obtain the column density measurements from the
\citet{savage_etal_03} catalog of $\sim100$ sight lines, 
which excludes high velocity \oxysix\ 
and is based on the \citet{wakker_etal_03} data analysis.
Their four sight lines nearest to our off-cloud sight line
are those toward
PG 1626+554 ($\log{N} = 14.25 \pm 0.09 \pm 0.08$, where the
first uncertainty is the combined statistical and continuum
placement 1 $\sigma$ error in the $\log$ of the column density; 
the second is the 1 $\sigma$ systematic error, which, in 
this case, is not 
necessarily correlated from one observation to the next), 
Mrk 487 ($\log{N} = 14.25 \pm 0.11 \pm 0.04$), 
Mrk 290 ($\log{N} = 14.21 \pm 0.12 \pm 0.05$),  
and Mrk 876 ($\log{N} = 14.43 \pm 0.02 \pm 0.05$).
The average Milky Way \oxysix\ column density on these 4
lines of sight is 
$1.97^{+0.21+0.14}_{-0.17-0.12}\times 10^{14}$~cm$^{-2}$.

The average angular separation
between these sight lines and our own is 7.3$\degr$ and
the maximum is $<10\degr$.   
The average deviation in the mean column density for
pairs of sight lines having \mbox{$\sim7\degr$} of separation
in the \citet{savage_etal_03} catalog (their figure 11) 
is 24$\%$.
%
%
Following the logic of \citet{shelton_etal_07},
the rms deviation between the mean of the parent population
and the mean of a sample of four observations drawn randomly
from the parent population should be 
$\sqrt{\pi}/4$ times as large as the average
deviation between 2 sight lines, thus 
$11\%$, and the 
intrinsic fluctuation between column densities along different
directions should be a factor 
$\sqrt{\pi}/2$ times the average deviation between
2 sight lines, thus 
$21\%$.
Combining these terms in quadrature yields 
$24\%$.   
This is larger than the average  
statistical plus continuum placement error
and the average systematic error obtained from 
the four individual column density measurements.
Here, we add the 3 sources of uncertainty in quadrature
in order to obtain an estimate of 
the uncertainty in the estimated column density for our
direction.
The column density and combined error are
$1.97^{+0.53}_{-0.51}\times 10^{14}$~cm$^{-2}$.

The column density of local material must be subtracted from
the line of sight column density.    The \citet{savage_lehner_06}
survey of \oxysix\ toward nearby white dwarfs includes 9
sight lines that are within 20$\degr$ of our off-cloud
direction.   Ions of \oxysix\ were 
not detected on 4 of these
sight lines.   In those cases, \citet{savage_lehner_06} listed
$2\sigma$ upper limits.   
We treat these cases as if the $1\sigma$ upper limits
were half of the $2\sigma$ upper limits and the detected
values were 0, so that we can find the average for the
9 sight lines.   The resulting average volume density of \oxysix\ is 
$1.5^{+1.2}_{-0.4} \times 10^{-8}$~cm$^{-3}$.
Although there are reports (i.e., \citealt{barstow_etal_09})
claiming that only a few of the previous \oxysix\ detections
for nearby sight lines can be shown to be interstellar, as 
opposed to photospheric or ambiguous (generally due to
the similarities between the observed 
velocities
and the velocities of photospheric lines in the spectra), 
a culled dataset
is not yet available.   Thus, we use the \citet{savage_lehner_06}
dataset.

We estimate the maximum 
extent of the Local Bubble from a survey of the Local Cavity's 
wall, in this case the
\naone\ and \catwo\ survey of \citet{welsh_etal_10}.   In
Figures 14 and 17 of that paper, the Local Cavity extends
75~pc in the direction of our observation.
Taking this and the average volume density yields an estimate
of the Local Bubble's column density of \oxysix\ 
of $3.4^{+2.8}_{-1.0} \times 10^{12}$~cm$^{-2}$.   
This is so small compared with the full column density
through the Galactic disk and halo that approximations
are inconsequential.   Subtracting this value from
the full column density yields a halo \oxysix\ column density of 
%
%
$N_{OVI} = 1.9 \pm 0.5 \times10^{14}$~cm$^{-2}$.
Again, the unrounded measurement
values are used in the calculations, but the rounded results
are reported.

\subsection{The Halo \oxyseven\ and \oxyeight\ Intensities}
\label{ovii_ovii_estimate}

\subsubsection{Intensity Measurements from 
\xmmnewton\ Observations}
\label{sect:xmm}

There are 33 archival \xmmnewton\ observations within 
5\degr\ of our \fuse\ pointings that have at least some 
EPIC-MOS exposure. All but 3 of the observations were 
badly affected by soft proton contamination. 
We have processed the data from these 3 observations using the method
described in \citet{henley_shelton_10}. We removed times affected by
soft proton flares, and also times when the solar wind proton flux
exceeded $2 \times 10^8$ cm$^{-2}$ s$^{-1}$ 
(the latter step was to reduce contamination
from solar wind charge exchange X-rays; see Section 2.4.2, below). We
extracted spectra from the blank sky regions of the EPIC-MOS chips,
and measured the intensities in the O VII K$\alpha$ 
triplet (569-574~eV)
and O VIII K$\alpha$ line, accounting for the effects of residual soft
proton contamination and
the extragalactic background (the spectrum of which we assumed to be
10.5($E$/keV)$^{-1.46}$~ph 
cm$^{-2}$ s$^{-1}$ sr$^{-1}$ keV$^{-1}$; 
\citealt{chen_etal_97}). 
For the portions of observation 0200750601 ($\ell = 90.0\degr, b = 38.4\degr$)
not affected by soft proton flares, the solar wind proton flux exceeded
our standard filtering threshold of $2 \times 10^8$ cm$^{-2}$ s$^{-1}$. 
However, the
proton flux was not unusually large, and did not exceed 
$2.8 \times 10^8$ cm$^{-2}$ s$^{-1}$. 
We therefore did not carry out proton flux filtering on this
observation. 
The X-ray intensity measurements are listed in
Table~\ref{table:ovii_oviii}.

\begin{deluxetable}{lccccc}
\tablewidth{0pt}
\tablecaption{\oxyseven\ and \oxyeight\ Intensities from
\xmmnewton\ Data}
\tablehead{
\colhead{Obs. ID } 
& \colhead{Obs. date }
& \colhead{$\ell$ }
& \colhead{$b$ }
& \colhead{$I_{OVII}$ }
& \colhead{$I_{OVIII}$ } \\
\colhead{ }     
& \colhead{ }
& \colhead{ }
& \colhead{ }
& \colhead{(LU)}
& \colhead{(LU)}
}
\startdata
0049540401 & 2003-08-19	& 86.567	& 40.932	& $11.6^{+0.8}_{-0.7}$	& $3.7^{+0.5}_{-0.4}$	\\

0200750601 & 2005-09-07	& 90.020	& 38.406	& $6.8^{+0.9}_{-0.6}$		& $1.0^{+0.4}_{-0.3}$	\\

0302310101 & 2004-09-29	& 86.462	& 41.115	& $15.3^{+0.8}_{-1.8}$	& $2.2^{+0.7}_{-0.5}$	\\

%
%
\enddata
\tablecomments{For these measurements, the extragalactic 
background component was removed.   During the
estimation of its strength, the background was assumed to have 
been partially absorbed by Galactic material, whose \hone\ column 
density was measured by the LAB survey. 
}
\label{table:ovii_oviii}
\end{deluxetable}

\pagebreak
\subsubsection{SWCX Abatement}

The measured \oxyseven\ and \oxyeight\ intensities 
differ from one sight line to the next by factors of 
$\stackrel{<}{\sim} 2$ and $\stackrel{<}{\sim}4$,
respectively, in spite of the fact that the
\rosat\ 3/4 keV map of this region
shows a relatively uniform surface brightness.    
The variation found between the three \xmmnewton\ observations
is not due to
absorbing material along the lines of sight, as the
obscuration toward
$\ell = 90.0\degr, b = 38.4\degr$ is insufficient
to explain the relative dimness of its \oxyseven\ measurement.
The variation is probably due to 
flares of contamination by solar wind charge exchange (SWCX) 
X-rays.   Such X-rays were generated after
solar wind ions charge exchanged with neutral gas 
in the heliosphere and Earth's extended atmosphere
\citep{robertson_cravens_03,koutroumpa_etal_06}.   
Observational work has shown that SWCX contamination can vary
significantly as a function of time
\citep{snowden_collier_kuntz_04,fujimoto_etal_07,henley_shelton_08,carter_sembay_10}, even during 
periods having modest solar wind proton fluxes
\citep{henley_shelton_10}.  
The intensity variation that we found among the 
three Draco pointings,
which were observed during different years
(see Table~\ref{table:ovii_oviii}),
is within the range of the variation
seen among multiple \xmmnewton\ 
observations of the same direction 
in the \citet{henley_shelton_10} catalog and is probably
due to variability in the SWCX X-ray intensity.
With this in mind, we take the
observation having the minimum oxygen intensities as that which
is least affected by time-variable SWCX X-rays, and
adopt these measurements.
Nonetheless, this observation, 
($\ell = 90.0\degr, b = 38.4\degr$, having
$I_{OVII} \sim 7$~LU and 
$I_{OVIII} \sim 1.0$~LU) 
is subject to low-level, slowly-varying SWCX.
Note that the pointing direction is only $1.2\degr$ from
our \fuse\ and \fimsspear\ direction and that
the \oxyseven\ and \oxyeight\ intensities
are not unusual compared with 
those seen along $b \stackrel{>}{\sim} 30\degr$ 
directions in
the \citet{henley_shelton_10} \xmmnewton\ catalog.

\subsubsection{The Intrinsic Halo \oxyseven\ and \oxyeight\
Intensities}
\label{subsect:intrinsicovi}

Low-level SWCX contamination is not the only
contribution to the observed spectra.
SWCX X-rays, together with Local
Bubble X-rays, if they exist, form a foreground
component,
$I_{fg}$,  which is not subject to interstellar 
extinction.
The emission from the Galactic halo, $I_{h}$, 
is subject to extinction having an optical depth of 
$\tau$ and combines with the foreground component as
$I_{obs} = I_{fg} + I_{h} \exp(-\tau)$ 
to produce the observed intensity,
$I_{obs}$.
The foreground component
can be estimated from shadowing observations of
nearby clouds
(see Table 4 in \citet{gupta_etal_09}) and SWCX
models (see Table 4 in \citet{koutroumpa_etal_07})
for cases not subject to strong SWCX flares.
Here, we take the foreground 
component to be
$I_{fg,OVII} = 1.5\pm1.0$ LU and 
$I_{fg,OVIII} = 0\pm1.0$ LU.
The optical depth was estimated in three different ways:
by using the weighted average of the \hone\ column densities
on nearby pointings in 
the LAB survey \citep{kalberla_etal_05}, 
using relations in \citet{snowden_etal_2000} to
convert the 100~$\mu$m intensity from the
\dirbe-corrected \iras\ maps \citep{schlegel_etal_98}
to $N_H$, and
using relations in \citet{guever_oezel_09} to convert
the reddening (E(B-V)) obtained from the
\dirbe-corrected \iras\ maps to $N_H$.   In each case,
the absorption cross sections of 
\citet{balucinskachurch_mccammon_92} were then used.
Table~\ref{table:halo_ovii_oviii} lists the data from
these surveys and the intrinsic 
\oxyseven\ and \oxyeight\ intensities calculated
by solving for $I_h$ in  
$I_{obs} = I_{fg} + I_{h} \exp(-\tau)$.
Here, we adopt the average values:
$I_{h,OVII} =  6.4^{+1.4}_{-1.5}$~LU and
$I_{h,OVIII} = 1.0 \pm 1.2$~LU.

\begin{deluxetable}{lccc}
\tablewidth{0pt}
\tablecaption{Intrinsic Halo \oxyseven\ and \oxyeight\ 
Intensities for $\ell = 90.0\degr, b = 38.4\degr$ direction
}
\tablehead{
\colhead{Method of Measuring} 
& \colhead{$N_H$ }
& \colhead{$I_{h,OVII}$ }
& \colhead{$I_{h,OVIII}$ } \\
\colhead{Obscuration}     
& \colhead{$10^{20}$~cm$^{-2}$}
& \colhead{(LU)}
& \colhead{(LU)}
}
\startdata
LAB Survey	&	$1.93$	& $6.3^{+1.6}_{-1.4}$	& $1.1 \pm 1.2$	\\

$I_{100\mu{\rm{m}}}$ (0.920 MJy/sr)	& $1.48$	& $6.2^{+1.3}_{-1.5}$	& $1.0 \pm 1.2$	\\

E(B-V) (0.0241 mag)	& $1.65$	& $6.5^{+1.4}_{-1.5}$	& $1.0\pm1.2$ \\

%
%

\enddata
\tablecomments{ 
The assumed optical depth affects the 
removal of the extragalactic background.
For consistency, the data were
reprocessed whenever the assumed optical depth was changed.
The absorption cross-sections are from 
\citet{balucinskachurch_mccammon_92} 
with an updated He cross-section from \citet{yan_etal_98},
calculated using \citet{anders_grevesse_89} abundances. \\
Line 2: the conversion relation in
\citet{snowden_etal_2000} was used to convert from
the \citet{schlegel_etal_98} $I_{100\mu{\rm{m}}}$ intensity 
to $N_H$. \\ 
Line 3: the conversion relation from 
\citet{guever_oezel_09} was used to convert 
the \citet{schlegel_etal_98} value of E(B-V) to $N_H$.}
\label{table:halo_ovii_oviii}
\end{deluxetable}

\section{Discussion}
\label{sect:discussion}

The ratio of $I_{OVI}$ to $N_{OVI}$ allows us to
estimate the electron density ($n_e$) and thermal pressure
($p_{th}$)
of the \oxysix-rich gas in the halo.
The calculation method is the same as that
used in \citet{shelton_etal_01} and \citet{dixon_etal_01}, but is
used here with the intrinsic intensity and
column density of the \oxysix\ ions located above the disk.
It yields
$n_e = 0.011\pm0.004$~K~cm$^{-3}$ 
$p_{th}/k = 6500^{+2500}_{-2600}$~K~cm$^{-3}$, 
where $k$ is Boltzmann's constant. 
This pressure estimate and that found for the
southern off-filament direction 
(7000 - 10,000~K cm$^{-3}$; \citealt{shelton_etal_07})
are the only pressure
estimates for extraplanar \oxysix-rich gas 
in which the local
contributions to the observed quantities were
subtracted before the pressure was calculated.
In addition, several estimates have been made without
the subtraction of the relatively small local contributions
to the \oxysix\ intensity and column density
(i.e., \citealt{shelton_etal_01}: $P_{th}/k \sim 5300$ to
$14,000$ K cm$^{-3}$, with the range reflecting the
uncertainty in the location of the \hone\ gas;
\citealt{dixon_etal_01}:  $P_{th}/k \sim 20,000$ K cm$^{-3}$
for $T = 10^{5.3}$~K gas;
\citealt{shelton_02}: $P_{th}/k \sim 3700$ to 4900 K cm$^{-3}$,
depending upon the location  of the \hone\ gas;
\citealt{dixon_sankrit_08}: $P_{th}/k \sim 5300$ to 7400 K cm$^{-3}$
for the Fairall 9 direction and 3000 to 3600 K cm$^{-3}$
for the NGC 625 direction, with the ranges 
reflecting the uncertainties in the locations of the \hone\ gas).

The thick-disk/halo \oxysix-rich 
gas is expected to reside between heights of
$|z| \sim 160$~pc (the height of the shadowing cloud 
that was used in the
observation of local \oxysix\ photons and discussed in 
Subsection~\ref{subsect:intrinsic}) and $\sim2300$~pc (the
scale height of \oxysix\ ions, \citet{savage_etal_03})
and has a much larger thermal pressure than had
been estimated for this region 
based on observations and
the assumption of hydrostatic equilibrium
($P_{th}/k < 2200$~K cm$^{-3}$; \citealt{ferriere_98}).   
The discrepancy could be due to a lack of hydrostatic
balance, as would occur if disruptive events occurred 
too frequently for the thick disk to re-relax.
The discrepancy could 
also be due to
the assumption that the gas is at its CIE temperature.  The calculated
pressure is proportional to temperature divided by a
function of temperature and has a minimum of 
$p_{th}/k = 3300 \pm 1300$~K~cm$^{-3}$
at $T = 9.6 \times 10^4$~K.
Alternatively, the discrepancy for our direction can be reduced by 
relaxing our assumption that the obscuring material
along the sight line resides below the extraplanar
\oxysix.   If, instead, it lies above, then
the extraplanar \oxysix's intensity would be
$3480^{+940}_{-1000}$~LU
(from Subsection~\ref{subsect:intrinsicovi}), implying    
a thermal pressure of
$p_{th}/k = 4700^{+1800}_{-1900}$~K~cm$^{-3}$.

The ratios of the intrinsic extraplanar intensities
of \carfour, \oxysix, \oxyseven, and
\oxyeight\ are useful tests for halo models.
Here we use them to test several models, beginning
with a model in which gas is heated to a high temperature
and then allowed to cool radiatively.   
This model is plausible, given that its predictions
have been found consistent
with the ratios of \oxysix, \oxyseven, and \oxyeight\
{\it{column densities}} observed for the
PKS 2155-304 sight line by \citet{heckman_etal_02}.
An application of the radiative cooling model involves gas that
was accreted from intergalactic space.  
Although \citet{henley_etal_10} show that an extended hot halo of 
accreted gas ($r \sim 10$s of kpc) may be too faint to account 
for the X-ray emission seen in the \xmmnewton\ band 
(including the \oxyseven\ and \oxyeight\ lines), the
model X-ray luminosities used by \citet{henley_etal_10} 
excluded emission from near the disk.    If accreted
material near the disk contributes to 
the observed O VII and O VIII emission then the model is
not ruled out.
Here we predict theoretical intensity ratios and compare
with our observations.

\citet{lei_etal_09} presented
a simple model in which radiative cooling of
hot, shocked intergalactic gas 
results in a temperature-stratified
layer of accreted gas.   
In such a model, the intensities in the  
\carfour, \oxysix, \oxyseven, and \oxyeight\ lines
can be calculated from
\begin{equation}\label{eq:intensity}
%
I = \left(\frac{k}{4 \pi}\right) \left(\frac{3}{2} + s \right) \left(\frac{1}{A} \frac{dN}{dt}\right)\int \frac{n_h}{n_{pi}} \frac{\epsilon(T)}{\Lambda_N(T)} dT ,
\end{equation}
where $n_h$ and $n_{pi}$ are the volume densities of hydrogens
and positive ions, respectively, $s$ is $0$ for isochorically 
cooling gas
\citep{lei_etal_09}
and $1$ for isobaricically cooling gas
\citep{edgar_chevalier_86}, 
$\frac{1}{A}\frac{dN}{dt}$ is the unknown rate at which particles 
are accreted per unit cross sectional area, 
$\epsilon(T)$ is the emission coefficient,
and $\Lambda_N(T)$ is the cooling function for radiatively
cooling gas.   Following \citet{sutherland_dopita_93},
$\Lambda_N$ has units of ergs cm$^{3}$ s$^{-1}$ and
is defined such that the energy loss per unit volume per
unit time equals $n_e n_{pi} \Lambda_N(T)$, where $n_e$ is the
volume density of electrons.
We take $\Lambda_N(T)$ from Table 6 in 
\citet{sutherland_dopita_93}, which approximates the 
ionization levels of the gas as those in CIE plasma.  
Solar abundances from \citet{anders_grevesse_89} are used.
We take the emission coefficients from the Xspec databases
for the Raymond \& Smith and APEC models for
radiation from optically thin thermal plasmas in collisional
ionizational equilibrium. 
The elemental abundances,
which are assumed to be solar from 
\citet{anders_grevesse_89}, and the fraction of atoms in
the relevant ionization stage are factored into these
coefficients.    
The emission coefficients 
have units of ergs cm$^{3}$ s$^{-1}$ and are defined such
that the energy loss per unit volume per unit time 
is $\epsilon(T)n_e n_H$.
Because $\Lambda_N(T)$ and
$\epsilon(T)$ are tabulated at discrete values of $T$, we 
approximate the intensity integral with a sum.

	Not knowing the value of $\frac{1}{A}\frac{dN}{dt}$,
we report only the ratios of the intensities for comparisons
with the observations.
The model ratios of 
$I_{CIV} : I_{OVI} : I_{OVII} : I_{OVIII}$ are
$0.13 : 1.00 : 0.055 : >0.26$ from the Raymond \& Smith 
coefficients and
$>0.068 : 1.00 : 0.057 : >0.26$ from the APEC coefficients.
Upper limit signs appear for ions whose
emissivity functions extend to higher or lower temperatures
than were covered in the tables.    (The Sutherland
\& Dopita tables run from $T = 10^4$ to $10^{8.5}$~K, 
while the Raymond \& Smith
tables extend from $T = 10^4$ to $10^8$~K and the
APEC tables extend from $T = 10^5$ to $10^{8.9}$~K.)
These ratios were calculated from intensities in units of 
photons sec$^{-1}$ cm$^{-2}$ sr$^{-1}$,
rather than intensities in 
units of ergs sec$^{-1}$ cm$^{-2}$ sr$^{-1}$.
The derived ratios have 
far too little \carfour\ and \oxysix\ flux,
or conversely, far too much \oxyseven\ and \oxyeight\ flux
for consistency 
with the observationally derived ratios 
$1.4\pm0.6 : 1.0 : 1.3\pm0.5\times 10^{-3} : 2.2^{+2.6}_{-2.5}\times 10^{-4}$
a problem that 
also plagues the 
extraplanar gas near the filament in the southern galactic
hemisphere ($I_{CIV} : I_{OVI} : I_{OVII} : I_{OVIII} =$ 
$1.0\pm0.4 : 1.00 : 1.3\pm0.2\times10^{-3} : 
3.4^{+0.6}_{-0.5}\times 10^{-4}$, 
from intensities presented in \citealt{lei_etal_09}).
This problem cannot be 
remedied by adjusting the gas phase metal abundances, although
it can be reduced by
decreasing the temperature of the upper limit of integration
in the intensity integral.
The temperature of
the freshly accreted gas should be around
$10^{6.5}$~K before it begins to cool \citep{lei_etal_09}.
Setting the upper limit of integration to 
$T = 10^{6.5}$~K results in
$I_{CIV} : I_{OVI} : I_{OVII} : I_{OVIII} =$ 
$0.13 : 1.00 : 0.037 : 0.031$ and
$>0.068 : 1.00 : 0.041 : 0.027$ for the Raymond \& Smith and
the APEC coefficients, respectively.     
The predicted ratios of $I_{OVII}$ to $I_{OVI}$ and 
$I_{OVIII}$ to $I_{OVI}$ are still far
too large.  
Our model could be tuned by 
further adjusting the upper temperature cut off; however, it would
be more logical 
to interpret this model as a source of 
some of the background
\oxyseven\ and \oxyeight, while the majority of the 
\carfour\ and 
\oxysix\ on the off-cloud sight line derive from other 
phenomena.

Localized, sporadic injections of energy
preferentially produce \carfour\ and \oxysix\
and could
explain the uneven distribution
of \oxysix\ ions across the sky (see \oxysix\ distributions in
\citealt{savage_etal_03,dixon_etal_06}).
Simulations of impulsive injections of energy in the form of
supernova explosions have been done for
explosion energies of $5 \times 10^{50}$~ergs
occurring in an ambient medium that has 
a combined magnetic and cosmic ray pressure of 
1800 K cm$^{-3}$ 
\citep{shelton_06}.   
These large injections of energy engender shocks that
heat and ionize the gas.   Although the early post-shock
temperatures are above $3 \times 10^6$~K, the expanding spherical
shockfronts weaken with time, reducing the post-shock
temperature.   As this happens, the gas inside the bubble 
cools by adiabatic expansion.   Later, it cools primarily 
by radiative cooling.   Each of these processes 
skews the temperature distribution in favor of 
\carfour\ and \oxysix-rich gas rather than
\oxyseven\ and \oxyeight-rich gas.
When averaged over time and space,
the resulting bubbles have ratios of 
median intensities of
$I_{CIV}:I_{OVI}:I_{OVII}:I_{OVIII} 
 = 0.39:1.0:3.4\times10^{-4}:3.7\times10^{-5}$
(assuming \citealt{anders_grevesse_89} abundances).
These ratios were
calculated from a Monte Carlo simulation
that included the luminosity of individual SNR bubbles
as a function
of time, variation in ISM conditions and therefore
SNR characteristics as a function of 
height above the plane, and the likelihood that 
a sight line would pass though any given SNR, based the 
SNR's size, height above the midplane, and the height distribution
of progenitor stars.  They, therefore,
represent the all high-latitude-sky average, which
is serendipitous as our observed
intensities are also near the average of those observed 
across the high-latitude sky.
In contrast with the accreting gas model, the 
impulsive injection model
predicts too little emission from \oxyseven\ and \oxyeight\ 
per photon emitted by \oxysix.

Another possible source of extraplanar \oxysix,
\oxyseven, and \oxyeight\ is
hot gas that has welled up from the Galactic disk.
In the hydrodynamical simulations of \citet{joung_maclow_06},
SN explosions
stir the interstellar medium and drive hot gas into the 
thick disk.    
These simulations, which modeled 80~Myr of ISM evolution,
were extended to 155~Myr for
\citet{henley_etal_10}. (Note that these times
include the time taken for the initial conditions 
to be eradicated.)   
From them and the Raymond
\& Smith spectral code,
\citet{henley_etal_10} created synthetic spectra.
The synthetic spectra from time periods after 140~Myr
were found to be
at least as bright in 0.4 to 2.0~keV photons
as the spectra extracted from \xmmnewton\ observations
after removing the local contribution.
Thus, such upwelled hot gas is a probable source of
\oxyseven\ and \oxyeight.   However,
model 
$I_{CIV}:I_{OVI}:I_{OVII}:I_{OVIII}$ ratios 
are not yet available for comparison with our observations.

Furthermore, there have been reports of enhanced X-ray 
intensities
associated with high velocity clouds, HVCs
\citep{hirth_etal_85,kerp_etal_99,bregman_etal_09}.
Our line of sight passes near the intermediate and high velocity
material associated with the Draco Nebula and Complex C, 
crosses some fast moving
material (see Section~\ref{subsect:oviintensity}), and
hypothetically could encounter materal left behind by 
an HVC, given that simulated HVCs 
shed material \citep{heitsch_putman_09}.
Although intensity predictions from 
simulations are not yet 
available, we can assume that several known
physical processes may be active.    Supersonic collisions
between high velocity clouds and thick disk/halo
gas would shock heat the gas.  The clouds may experience
turbulent mixing and evaporation if they are passing
through a hot ambient medium.  Radiative cooling will
also operate.  \citet{slavin_etal_93} analytically calculated the
\carfour\ and \oxysix\ intensities resulting from model 
turbulent mixing layers.  The $I_{CIV}:I_{OVI}$ ratios
from their predictions ranged from 
1.7 to 16, 
depending upon the assumed conditions.
The low end of the predicted $I_{CIV}:I_{OVI}$ ratio
is within the $1\sigma$ error bars of the observationally
determined ratio and is from models having 
mixed gas temperatures of $\bar{T} = 10^5$~K.
\citet{kwak_shelton_10} hydrodynamically simulated
mixing layers and calculated the quantities of \carfour, \nitfive,
and \oxysix, ions in a
time dependent manner (i.e., not assuming CIE).
Subsequently, they calculated the
ratio of the \carfour\ photon intensity to the \oxysix\ 
photon intensity for their primary model, 
Model A, and for a similar model with 10 times as much thermal 
pressure after 50~Myr of evolution, finding it to be 
2.4 and 4.9, respectively (K. Kwak, personal communcation, 2010).
Not only are the ratios of \carfour\ to \oxysix\ intensity
larger in turbulent mixing layer models than in 
radiative cooling and supernova remnant models, but the
ratios of \carfour\ to \oxysix\ column densities are
also larger
\citep{slavin_etal_93,esquivel_etal_06,kwak_shelton_10}.
None of these papers predicts the intensities of
\oxyseven\ and \oxyeight\ and such values would
depend strongly upon the assumed depth of the hot reservoir.
However, we can assume that the greater the number of
mixing layers, the greater the $I_{OVI}:I_{OVII}$
and $I_{OVI}:I_{OVIII}$ ratios will be.

In the foregoing analyses, we have ignored scattering of
UV background
photons by interstellar
\carfour\ and \oxysix\ ions because we expect the scattered
intensities to be small.
\citet{shelton_etal_01} estimated the
intensity of photons scattered by \oxysix\ ions, 
finding it to be much less than (i.e., $6\%$ of)
the intensity produced by line emission for
their Case 3b, which is similar to our case in that
it assumed that most of the \oxysix\ photons originated
in the halo and that there is some line-of-sight
obscuration.
By similar logic, we expect that scattering by 
\oxysix\ ions has
negligibly affected our observations. 
To examine the magnitude of scattering by \carfour\ ions 
in comparison with that by \oxysix\
ions, we use equation 14 from \citet{shelton_etal_01} 
for the scattered intensity, 
$I_{sco} = \frac{a2b}{\lambda} I_{sci} F(\tau_o)$.
The background continuum specific intensity, $I_{sci}$,
near 1550 \AA\ is 
$4.1 \times 10^{-21}$ 
ergs cm$^{-2}$ s$^{-1}$ sr$^{-1}$ Hz$^{-1}$
\citep{bowyer_91}, which 
is only 1/10 of the intensity near 1030 \AA,
$4.4 \times 10^{-20}$
ergs cm$^{-2}$ s$^{-1}$ sr$^{-1}$ Hz$^{-1}$
(\citealt{mathis_etal_83} and
assumed in the \citet{shelton_etal_01} analysis).
If the line profile is thermally broadened and the
gas is in CIE, then
the velocity spread parameter, $b$, of \carfour\
is $2/3$ of that of \oxysix.
Its wavelength,
$\lambda$, is obviously larger.  
We expect the column density of \carfour\
ions to be around $60\%$ of that of \oxysix\ ions, given
the ensemble statistics reported in \citet{savage_wakker_09}.
Thus, the optical depth of scatterers, $\tau_o$, will be
smaller, and $F$, which is a function of $\tau_o$, will be
smaller for \carfour\ than for \oxysix.  
The variable $a$ is related to absorption by
intervening dust and is $6\%$ larger at 1550 \AA\
than at 1030 \AA.  As a result of these factors, the intensity 
scattered by \carfour\ ions is even smaller than that
scattered by \oxysix\ ions.


\section{Summary and Conclusions}\label{sect:summary}

In this paper, we discuss measurements of
the \carfour, \oxysix, \oxyseven,
and \oxyeight\ intensities.   The \oxysix, \oxyseven,
and \oxyeight\ data are newly presented
here, while the \carfour\ intensity is taken from
\citet{park_etal_09}.     
Each of the measured intensities was taken from 
$\ell = 90\degr, b \sim40\degr$, a region that
is near to but not toward
the Draco cloud, and is fairly typical of high latitude
values.

We have subtracted
the contributions made by local diffuse gas and
compensated for the absorption loss due to intervening material,
in order to obtain the intrinsic intensities of photons
that originated above the plane.
While this region is often called
the halo, especially when discussing the diffuse X-ray 
background, most of the observed ions reside in the thick 
disk.   We find a high pressure 
$p_{th}/k = 6500^{+2500}_{-2600}$~K~cm$^{-3}$, 
see Section~\ref{sect:discussion}
for the \oxysix-rich gas in this region.
It is consistent with the measurement of extraplanar
\oxysix\ made in the southern Galactic hemisphere
and is greater than expected from static models of the thick
disk.   Such a high pressure could result from the thick
disk being disturbed on a faster timescale than the relaxation
timescale.   
In addition, 
if the true temperature of the gas is much less than the CIE temperature,
then the true pressure may be somewhat less than that derived from
our measurements.

Each of our ions traces hot gas, although they
sample different, sometimes overlapping temperature subregimes 
(\carfour\ traces $\sim1 \times 10^5$~K gas, 
\oxysix\ traces $\sim3 \times 10^5$~K gas,
\oxyseven\ traces $1 \times 10^6$~K gas, and
\oxyeight\ traces $\sim3 \times 10^6$~K gas
if the gas is in collisional ionizational equilibrium).
The hot gas in these subregimes is thought to be causally and 
sometimes physically related.   One relationship occurs 
through radiative
cooling, because gas that was heated to a few million degrees 
will cool through the other subregimes if left undisturbed.
Another relationship occurs through energy transfer, either
by thermal conduction or by mixing.   If very hot gas
abuts significantly cooler gas, then
a \carfour\ and/or \oxysix-rich zone develops between the
hotter (\oxyseven\ and/or \oxyeight-rich) gas and the cooler
gas.


The ratio of intensities provides
a diagnostic of the heating mechanism.   Mechanisms
that heat gas to several times $10^5$~K favor 
\carfour\ and \oxysix, while mechanisms that heat the gas
to higher temperatures boost the relative \oxyseven\ and
\oxyeight\ intensities.    It should be noted
that atomic physics favors the resonance lines of lithium-like 
ions (i.e., \oxysix\ 1032, 1038 \AA\ and \carfour\ 
1548, 1551 \AA\ emission) over the K $\alpha$ lines of
helium-like and hydrogen-like ions 
(i.e., the \oxyseven\ ``triplet''
between 561 and 574~eV and the \oxyeight\ 653 eV line),
if all other factors, such as abundance, are equivalent.
Therefore, when we indicate that heating the gas
to very high temperatures favors \oxyseven\ and
\oxyeight\ emission, we mean that $I_{OVII}/I_{OVI}$
and $I_{OVIII}/I_{OVI}$ are greater than in other models.
We do not mean to imply that our
\oxyseven\ and \oxyeight\ intensities 
are brighter than our
\oxysix\ intensity, because they are not.

The \oxysix\ intensity is the reference point for all of
our comparisons.
Our intrinsic, halo \oxysix\ intensity is 
$4770^{+1300}_{-1380}$~LU
(see Subsection~\ref{subsect:intrinsicovi})
and the 
$I_{CIV}:I_{OVI}:I_{OVII}:I_{OVIII}$ ratios are
$1.4\pm0.6 : 1.0 : 1.3\pm0.5\times 10^{-3} : 2.2^{+2.6}_{-2.5}\times 10^{-4}$
(see Section~\ref{sect:discussion}).
For the moment, we ignore \carfour, which we will return
to later.  The $I_{OVI}:I_{OVII}:I_{OVIII}$ ratios lie 
between the extremes formed
by two models, one of which is relatively preferential to
\oxyseven\ and \oxyeight\ while the other is relatively
preferential to \carfour\ and \oxysix. 
The first model is fairly simple.   It heats gas to a high
temperature ($10^{6.5}$~K), then allows it to cool radiatively.
This model, which may correspond to halo gas 
resulting from a smooth 
accretion of intergalactic gas, produces ratios of 
$0.13 : 1.00 : 0.037 : 0.031$ (assuming Raymond \& Smith
coefficients; the ratios found from APEC coefficients are
also presented in Section~\ref{sect:discussion}).
The second model 
was made from simulations of the bubbles blown by 
sporadic and isolated injections of energy.  The particular
model employed was for bubbles blown by supernova explosions.
The model ratios take into account the different numbers
of photons produced at different stages of the bubble's
life, including the short-lived early period
when the shock-front heats the gas
to temperatures well above $3 \times 10^6$~K, the
longer-lived stage when the shock-front 
heats the gas to a few times $10^5$~K, and the final 
much longer period when the SNR bubble radiatively cools.
Thus, this model includes the effects of injecting
energy to create a wide range of gas temperatures, in addition
to radiative cooling.
It produces ratios of 
$0.39:1.0:3.4\times10^{-4}:3.7\times10^{-5}$ 
(see Section~\ref{sect:discussion}).  
The observed $I_{OVI}:I_{OVII}:I_{OVIII}$ ratios are 
between the two comparison models.

We can conclude that in addition to some very hot gas
($T_{CIE} {\sim} 3\times 10^6$~K),
a broad temperature profile is needed.   
Not only could a broad 
temperature profile result from supernova bubbles, but
it may also result from
mixing or thermal conduction between very hot gas and 
cooler gas, 
injections of moderately hot fountain gas into the ``halo'', 
or other phenomena.

Interestingly, neither model produced more \carfour\
photons than \oxysix\ photons, yet the observations show
that \carfour\ is at least as intense as \oxysix\ in
both the Draco region and the southern filament region
(see \citealt{shelton_etal_07}).\footnote{Note that 
substantial $I_{CIV}:I_{OVI}$ ratios were also
reported in Table 1 of \citet{welsh_etal_07}.   Their
$I_{OVI}$ and $I_{CIV}$ were measured from \fimsspear\ 
observations directed in the north galactic polar region 
and supplemented by an additional measurement of the
north ecliptic pole made by \citet{korpela_etal_06}.
The extreme range in $I_{CIV}:I_{OVI}$ from the various
fields in which they detected both \carfour\ and \oxysix\
photons is $0.37\pm0.13$ to $1.0\pm0.11$.}
Adjusting the
C:O abundance ratio cannot solve this problem, as it would
require raising it from 
0.4 
\citep{anders_grevesse_89} 
to 1.6, which is unrealistic.
Instead, the large $I_{CIV}:I_{OVI}$ ratio indicates that
there exists a strong mechanism for producing \carfour\
photons.   Turbulent mixing between hot and cool gas is one
such mechanism.

\vspace{1cm}
\noindent Acknowledgments \\

We thank S. J. Lei for giving us the emission
coefficients used in Lei et al. 2009, 
K. Kwak for suggestions regarding high velocity clouds,
and Yangsen Yao for refereeing the manuscript.
RLS acknowledges
funding from NASA grant NNX07AH29G, awarded through the
\fuse\ guest investigator program. 
DBH acknowledges funding from NASA grant NNX08AJ47G,
awarded through the Astrophysics Data Analysis Program. 
This paper utilized observations obtained by the NASA-CNES-CSA 
{\it Far Ultraviolet Spectroscopic Explorer (FUSE)} 
mission operated by Johns
Hopkins University, supported by NASA contract NAS5-32985,
and observations obtained by \xmmnewton\, an ESA science
mission with instruments and contributions directly funded
by ESA Member States and NASA.


\end{document}